\documentclass[aps,prd,superscriptaddress,nofootinbib,amsmath,amssymb,amsfonts,floatfix,showpacs]{revtex4-1}

\usepackage[dvipdfmx]{graphicx}
\usepackage{amssymb}
\allowdisplaybreaks[4]
\usepackage{afterpage,dblfloatfix}
\usepackage{color}
\usepackage{xcolor}
\usepackage[colorlinks=true, allcolors=blue!75!black]{hyperref}
\usepackage{makecell}
\usepackage{multirow}
\usepackage{slashed}
\usepackage{wrapfig}
\usepackage{placeins}
\usepackage{ulem}
\usepackage{bm}
\usepackage[nameinlink]{cleveref}
\usepackage[caption=false]{subfig}
\usepackage{lineno}
\usepackage{flushend}

\begin{document}

\title{Technically Natural Suppression of Fifth Force}

\author{Kensuke Homma}
\email{khomma@hiroshima-u.ac.jp}
\affiliation{Graduate School of Advanced Science and Engineering, Hiroshima University, Kagamiyama, Higashi-Hiroshima, Hiroshima 739-8526, Japan}
\affiliation{International Center for Quantum-field Measurement Systems for Studies of the Universe and Particles (QUP), KEK, Tsukuba, Ibaraki 305-0801, Japan}
\author{Taishi Katsuragawa}
\email{taishi@ccnu.edu.cn}
\affiliation{Institute of Astrophysics, Central China Normal University, Wuhan 430079, China}
\affiliation{Laboratory for Compact Object Astrophysics and Astronomical Technology, Central China Normal University, Wuhan 430079, China}
\author{Shinya Matsuzaki}
\email{synya@jlu.edu.cn}
\affiliation{Center for Theoretical Physics and College of Physics, Jilin University, Changchun, 130012, China}

\date{\today}

\begin{abstract}
Light scalars generically mediate a fifth force incompatible with local tests of gravity unless their couplings are parametrically suppressed or screening mechanisms are introduced. We demonstrate that such suppression can arise from symmetry. We propose a $Z_2$-symmetric mirror extension of the Standard Model within a bi-conformal gravity construction, where spontaneous breaking of scale invariance produces a light scalaron as a pseudo-Nambu-Goldstone boson. This scalaron couples to the difference of trace anomalies between the Standard Model and mirror sectors. 
We find a parameter-independent correlation between the fifth-force strength $\alpha$ and the scalaron mass $m_\sigma$, with the proportionality set by QCD observables and the electroweak scale. 
The Standard Model predicts $\alpha \lesssim 10^{-3}$ above 10~cm scales for $m_\sigma \lesssim 10^{-6}$ eV, which is directly in the target window of next-generation experiments.
In contrast to environmental screening mechanisms, this suppression mechanism follows directly from symmetry rather than nonlinear scalar dynamics.
\end{abstract}

\maketitle

\section{Introduction}

Light scalar fields coupled to matter arise in many contexts beyond the Standard Model (SM) and general relativity. 
They appear as dark energy candidates (such as scalars in modified gravity or quintessence models), light dark matter (dilatons, axionlike particles), and moduli in string theory. 
A generic prediction of such fields is a long-range ``fifth force'' mediated by the scalar, which has experimentally been constrained to be extremely weak~\cite{Adelberger:2009zz}. 
To reconcile such predictions with the stringent experimental constraints, various strategies have been developed to suppress or ``screen'' the scalar coupling to the SM sector.

From perspectives of modified gravity, phenomenological screening mechanisms (chameleon \cite{Khoury:2003aq}, symmetron \cite{Hinterbichler:2010es}, Vainshtein \cite{Vainshtein:1972sx}) offer possible escapes, which typically rely on environmental dependencies or nonlinear dynamics rather than on fundamental particle physics principles. 
For instance, the chameleon mechanism suppresses long-range fifth forces in local observations; the scalar-field mass depends on the environment, and the scalar field becomes heavier as the ambient matter density is larger. 
However, since these theoretical frameworks generally lack an underlying symmetry to protect their couplings, screening mechanisms often require specific forms of the scalar potential or fine-tuned parameters to achieve the necessary suppression.

The pioneering work of Fujii \cite{Fujii:2003pa, Fujii:2015iuz} proposed a light scalaron, and its mass was shown to be generated radiatively via the conformal anomaly.
By adopting classical scale invariance, this formulation strongly constrains the form of the action and ensures that all dimensionful parameters are generated dynamically through quantum effects or spontaneous symmetry breaking. 
However, in the original model, the scalaron ($\sigma$) is universally coupled to the trace of the energy-momentum tensor in the SM.  
This scalar exchange contribution to the nucleon probe gives a conventional dilatonic fifth force with the Yukawa-type potential 
$V(r) \sim  G_N m_N^2 e^{- m_\sigma r}/(4\pi r)$~\cite{Fujii:2003pa, Fujii:2015iuz} (and see also, e.g., \cite{Kaplan:2000hh}). 
Due to the absence of an intrinsic screening mechanism, fifth force inevitably reaches orders of magnitude above experimental bounds.
To make this elegant framework phenomenologically viable, it is essential to suppress the fifth force via an underlying symmetry rather than relying on phenomenological adjustments, while preserving its symmetry-based structure.

In this paper, we present a fundamentally different approach: a new screening mechanism rooted in symmetry, anomaly structure, and vacuum alignment, which is technically natural~\cite{tHooft:1979rat, Wells:2013tta}. We extend Fujii's original framework by introducing a $Z_2$-symmetric mirror QCD sector embedded in a bi-conformal gravity construction. The scalaron $\sigma$ couples to the \textit{difference} of the trace anomalies of the QCD and mirror QCD sectors. In the $Z_2$-symmetric limit, $\sigma$ is an exactly massless Nambu-Goldstone boson at the vacuum $\sigma=0$. In this limit, the $Z_2$ symmetry forces the nucleon to couple equally to both sectors, canceling the dominant linear fifth force. When $Z_2$ breaking is included predominantly from the quark mass differences between the two sectors, QCD and mirror QCD confinement stabilize the scalaron $\sigma$ at a nontrivial vacuum, where $\sigma$ acquires the mass $m_\sigma \sim 10^{-6}$ eV via a finite two-loop mechanism akin to Fujii's original framework.
In this realistic $Z_2$ broken case, the linear fifth force is thus recovered, which is, however, shown to be smaller by a factor of $\sim {\cal O}(10^{-4})$ than the conventional dilatonic case. 
This residual force is determined by the well-measured QCD sigma terms. For the SM benchmark, the mechanism predicts a force strength $\alpha \lesssim 10^{-3}$, for the Yukawa potential $V(r) = -G_N m_N^2 (\alpha/r) e^{-m_\sigma r}$, above 10~cm scales for $m_\sigma \lesssim 10^{-6}$ eV, precisely in the window between current constraints and next-generation experimental sensitivity.

\section{Bi-conformal Mirror model}

We extend the SM by introducing 
 a mirror (``dark'') QCD sector with gauge group $SU(3)_D$ and three light quark flavors $u_D, d_D, s_D$, along with two real scalar fields $\phi$ and $\phi_D$. 
 The theory respects a $Z_2$ symmetry under which $\phi \leftrightarrow \phi_D$, SM quarks $q_{L,R} \leftrightarrow$ mirror quarks $q_{D L,R}$, and QCD gluons $G_\mu^a \leftrightarrow$ mirror gluons $G_{D\mu}^a$. 
 The SM Higgs field $\Phi$ is $Z_2$-even.

The $Z_2$ symmetry ensures that at high energies, the SM and mirror QCD sectors are identical, with equal gauge couplings $g_s(\mu) = g_{sD}(\mu)$ at some ultraviolet (UV) scale. The sole explicit breaking of the $Z_2$ symmetry originates from the electroweak Higgs-induced masses of the SM quarks. The mirror quarks, on the other hand, are assumed to remain (almost) massless, which can keep the mirror pion masses well below those of SM pions (see Appendix~\ref{sec:mirror_pions}).

The choice of a minimal $SU(3)_D$ sector with three light flavors ($u_D, d_D, s_D$) is the most predictive construction. Three flavors provide the necessary non-perturbative structures—a trace anomaly and chiral symmetry breaking—while mirroring QCD's successful low-energy dynamics. Fewer flavors would lack the strange-quark contribution essential for the quantitative prediction of $Z_2$-breaking effects, whereas a larger mirror sector, containing additional flavors or a full mirror electroweak sector, would introduce heavy states that decouple from the low-energy alignment mechanism without altering our core prediction (see Appendix~\ref{sec:mirror_generality}).

We embed the SM and mirror sectors in separate Jordan frames with the metrics $g$ and $g_D$ by a bi-conformal construction.
Based on the scale invariance and the $Z_2$ symmetry defined above, the model Lagrangian terms thus take the form:
\begin{align}
\begin{split}
    \mathcal{L}_J^{\text{SM,$\phi$}} 
    &= \sqrt{-g} \Big[ \frac{1}{2}\xi\phi^2 R(g) 
    + \frac{1}{2}(\partial_\mu \phi)^2 
    + \xi_h |\Phi|^2 R(g) + |D_\mu\Phi|^2 
    - \lambda_h|\Phi|^4 
    + \mathcal{L}_{\text{SM rest}}
    \Big], 
\\
    \mathcal{L}_J^{\text{Mirror,$\phi_D$}} 
    &= \sqrt{-g_D} \Big[ \frac{1}{2}\xi\phi_D^2 R_D(g_D) + \frac{1}{2}(\partial_\mu \phi_D)^2 
    + \mathcal{L}_{\text{Mirror QCD}} 
    \Big]. 
\label{eq:L_J_Mirror}
\end{split}
\end{align}
The scale invariance allows nonminimal couplings of scalars, $\phi$, $\phi_D$, and the SM Higgs field $\Phi$, to the Ricci scalar $R(g)$ with the coupling strengths 
$\xi_\phi = \xi_{\phi_D} = \xi$ and $\xi_h$, respectively, in which the former couplings are ensured by the $Z_2$ symmetry to be identical each other. The physics of this nonminimal couplings are out of our current scope. 
The mirror QCD sector introduced here is a specific realization of the broader ``mirror world''; e.g., see \cite{Berezhiani:2003xm} for a representative modeling. 
The bi-conformal construction involving the gravitational sector with the scale invariance is, however, substantially intrinsic to the current model setup. 
In Eq.(\ref{eq:L_J_Mirror}), $\mathcal{L}_{\text{SM rest}}$ includes the SM sector terms other than the Higgs sector, and $\mathcal{L}_{\text{Mirror QCD}}$ contains the $SU(3)_D$ gauge sector with three light quark flavors.

The scale invariance is spontaneously broken by the $Z_2$-symmetric vacuum expectation values (VEVs)
$\langle\phi\rangle = \langle\phi_D\rangle = M_P/\sqrt{\xi}$, generating the Planck scale.
These VEVs generate the Planck mass via the non-minimal couplings.
The $Z_2$ symmetry permits a Higgs portal coupling $\lambda_{\text{mix}}|\Phi|^2(\phi^2+\phi_D^2)$. We set $\lambda_{\text{mix}} = 0$ in what follows. This choice is technically natural: setting $\lambda_{\text{mix}} = 0$ enhances the symmetry of the classical action, and the $Z_2$ symmetry forbids radiative generation of this coupling. 


The two Jordan frames are mapped to a common Einstein frame metric $g_{\mu\nu}^E$ via separate Weyl transformations. Ignoring the $\xi_h$ term for the SM Higgs, for simplicity, the transformations go like
\begin{align}
\begin{split}
    g_{\mu\nu}^E &= \Omega^2 g_{\mu\nu}
    , \quad 
    \Omega^2 = \frac{\xi\phi^2}{M_P^2}
    , \\
    g_{\mu\nu}^E &= \Omega_D^2 g_{D\mu\nu}
    , \quad 
    \Omega_D^2 = \frac{\xi\phi_D^2}{M_P^2}
    .
\end{split}
\end{align}
The bi-conformal constraint $\Omega^2 g_{\mu\nu} = \Omega_D^2 g_{D\mu\nu}$ relates the Jordan-frame metrics, which ensures that both sectors describe the same physical spacetime.
Consequently, $g_{D\mu\nu}$ is not an independent gravitational degree of freedom, but is defined via the bi-conformal constraint. This constraint ensures that both sectors share a common Einstein-frame metric $g_{\mu\nu}^E$ while preserving independent Weyl scalings,
which reduces the physical metric degrees of freedom to the standard two (the Einstein-frame metric and the lightest scalaron fluctuation, dubbed $\sigma_-$, where the heavier one $\sigma_+$ is decoupled at the Planck scale, as clarified right below).

The $\phi$ and $\phi_D$ fields are decomposed into $Z_2$-even and odd eigenstates $\sigma_+ = (\phi + \phi_D)/\sqrt{2}$ and $\sigma_- = (\phi - \phi_D)/\sqrt{2}$. The $\sigma_-$ then yields null VEV, $\langle \sigma_- \rangle =0$, protected by the $Z_2$ symmetry and scale invariance, hence acts as a Nambu-Goldstone boson field as a consequence of the bi-conformal construction, while $\sigma_+$ is allowed to develop the VEV $\langle \sigma_+ \rangle = \sqrt{2} M_P/\sqrt{\xi}$.
Generating and stabilizing this VEV requires a potential — or an equivalent UV mechanism. The minimal assumption is a mass term $V(\sigma_+) = \frac{1}{2}m_+^2 \sigma_+^2$ with $m_+ \sim M_P$, which is technically natural (no symmetry protects $m_+$).

In contrast, $\sigma_-$ remains massless at the classical level. The $Z_2$ symmetry and the bi-conformal construction force the potential for $\sigma$ to be exactly flat. Radiative corrections will generate a potential, but these arise only at two-loop order due to the $\epsilon$ factor in the anomaly coupling (see Sec.~\ref{sec:mass_detail}). The resulting $\sigma$ mass is therefore determined entirely by a finite two-loop mechanism described in Sec.~\ref{sec:sigma-mass-gene}, not by unknown UV physics.

Thus, low-energy physics is governed by the $Z_2$-odd counterpart $\sigma_-$, which we shall rename $\sigma \equiv \sigma_-$.
Crucially, the low-energy physics of $\sigma$, including the fifth-force prediction, is completely insensitive to $m_+$ and the detailed form of $V(\sigma_+)$, as these degrees of freedom decouple below $M_P$.

\section{Particle Physics Screening via Vacuum Alignment}
\label{sec:vac} 

The bi-conformal geometry leads to a universal coupling of $\sigma$ to matter. Under a local scale transformation with parameter $\alpha(x)$, the $Z_2$ symmetry dictates opposite transformations for the two sectors: the SM Jordan frame metric $g$ and fields transform with $\alpha(x)$, while the mirror sector transforms with $-\alpha(x)$. After transforming to the Einstein frame, this yields the scale anomaly-induced couplings to matter (see Appendix~\ref{sec:geometric_derivation}):
\begin{align}
    \mathcal{L}_{\sigma T_\mu^\mu} = \frac{\sigma}{\sqrt{2}f_\Sigma} \left[ T^{\mu}_{\mu}(\text{SM}) - T^{\mu}_{\mu}(D) \right]
    \,,  
    \label{eq:anomaly_coupling_main}
\end{align}
where $f_\Sigma = \sqrt{6+1/\xi}~M_P$ is the scalaron decay constant, which is determined by performing the canonical normalization of $\sigma$ (see Appendix~\ref{sec:geometric_derivation}). 
This is an operator relation valid at all scales, independent of any dynamics. 
Thus, $\sigma$ couples to the {\it difference} of two traces of energy momentum tensors, $T_\mu^\mu({\rm SM})$ and $T_\mu^\mu(D)$. 

The operator $T^\mu_\mu$ in Eq.~\eqref{eq:anomaly_coupling_main} generically receives contributions from both perturbative and non-perturbative physics. The perturbative part, including the Higgs condensate contribution $\sim m_h^2 v_{\text{EW}}^2$ and short-distance gauge boson loops, is subtracted in the definition of the physical low-energy vacuum. This is the standard procedure in QCD sum rules and the operator product expansion: the physical condensates are defined as the difference between the full non-perturbative expectation value and the perturbative one. Consequently, only the non-perturbative parts arising from confinement survive in the low-energy effective theory relevant to the present-day $\sigma$.

Thus, the relevant quantities for the $\sigma$ potential are the non-perturbative condensates:
\begin{align}
\langle T^\mu_\mu(\text{QCD}) \rangle_{\text{NP}} \sim \Lambda_{\text{QCD}}^4, \quad
\langle T^\mu_\mu(D) \rangle_{\text{NP}} \sim \Lambda_D^4.
\label{conf}
\end{align}
Under a constant shift $\sigma \to \sigma + c$, the confinement scales transform oppositely due to the signs in Eq.~\eqref{eq:anomaly_coupling_main}: $\Lambda_{\text{QCD}} \to \Lambda_{\text{QCD}} e^{-c/(\sqrt{2}f_\Sigma)}$, $\Lambda_D \to \Lambda_D e^{+c/(\sqrt{2}f_\Sigma)}$. Since the non-perturbative vacuum energy density of each sector scales as the fourth power of its confinement scale, as in Eq.(\ref{conf}), the total effective potential generated by confinement thus goes like
\begin{align}
V_{\text{align}}(\sigma) = \frac{\kappa}{4} \Lambda_\text{QCD}^4 e^{-2\sqrt{2}\sigma/f_\Sigma} + \frac{\kappa}{4} \Lambda_D^4 e^{+2\sqrt{2}\sigma/f_\Sigma},
\label{eq:valign}
\end{align}
where $\kappa$ is a coefficient determined by non-perturbative QCD dynamics with a dilatonic background. (When we simply refer to the vacuum energy scale of QCD $\sim (1 \,{\rm GeV})^4$ together with $\Lambda_{\rm QCD} \sim 0.3$ GeV, the size of $\kappa$ is expected to be of ${\cal O}(10)$.) 
The form of the $\sigma$ potential has uniquely determined, up to the overall coefficient of $\kappa$, by the nonperturbative scale anomaly as a consequence of quantum field theory of QCD-like dynamics: the $\kappa$ constant can be set as universal for both QCD and dark QCD because a possible size of difference between the intrinsic scales for QCD and dark QCD can be absorbed into the definition of $\Lambda_{\rm QCD}$ and $\Lambda_D$. 
This potential, violating the scale invariance, deforms the original flat direction to induce \textit{vacuum alignment}.
There, the scalaron VEV $\langle\sigma\rangle$ develops at the nontrivial vacuum generated by the $Z_2$ breaking difference of $\Lambda_{\rm QCD}$ and $\Lambda_D$ as
\begin{align}
\frac{\langle\sigma\rangle}{\sqrt{2}f_\Sigma} = \frac{1}{2} \ln\left(\frac{\Lambda_{\text{QCD}}}{\Lambda_D}\right).
\label{eq:sigma_min}
\end{align}

\section{The scalaron mass generation}  
\label{sec:sigma-mass-gene} 

After the vacuum alignment described in Sec.~\ref{sec:vac}, the field $\sigma$ sits at its minimum $\langle \sigma \rangle$ determined by Eq.~\eqref{eq:sigma_min}. In this true vacuum, the SM particles have their usual masses from electroweak symmetry breaking, and the mirror QCD sector has its confinement scale $\Lambda_D$. When expanded around the true vacuum, any particle that acquires mass through scale symmetry breaking generically contributes to $m_\sigma^2$. As is clarified in detail in the Appendix~\ref{sec:mass_detail}, the dominant contributions arise at the two-loop level analogously to the Fujii original proposal~\cite{Fujii:2003pa, Fujii:2015iuz}, which is finite and takes the following universal form:
\begin{align}
m_\sigma^2 = \frac{{\cal C}}{(16\pi^2)^2} \frac{{\cal M}^4}{f_\Sigma^2}.
\label{eq:general_scaling_main}
\end{align}
${\cal M}$ is the characteristic mass scale of particles in the loop, and $\mathcal{C}$ aggregates all group theory factors and coupling constants. The factor $(16\pi^2)^{-2}$ is the characteristic two-loop suppression.

In the SM, the dominant contributions come from the electroweak scale, where the top quark provides the largest effect due to its large Yukawa and QCD couplings, which yield ${\cal M} = m_t$ and ${\cal C}\approx 196$ (see Appendix~\ref{sec:mass_detail} for a parametric estimate). The relevant two-loop contributions thus feed the physical scalaron mass as
\begin{align}
m_\sigma|_{\rm SM} \approx \frac{2.7 \times 10^{-6}\text{ eV}}{\sqrt{6+1/\xi}}.
\label{eq:m_sigma_SM}
\end{align}

\section{Fifth-Force Suppression}
\label{sec:5th-sup}

\subsection{$Z_2$ symmetric limit}

The $Z_2$ symmetry is approximately realized when the two QCD sectors have identical light quark features. In particular, in the massless quark limit, 
the theory is invariant under the transformation $\sigma \to -\sigma$ together with the exchange of the SM and mirror sectors. 
As noted around Eq.(\ref{eq:valign}), the vacuum of the theory in this $Z_2$ invariant limit is realized with respect to the background $\sigma$ field at $\sigma=0$. 
On the other hand, the Feynman-Hellmann theorem relates the derivative to the nucleon matrix element of the operator to which the constant background $\sigma$ field couples: 
\begin{align}
    \frac{\partial m_N(\sigma)}{\partial (\sigma/\sqrt{2}f_\Sigma)} 
    = 
    \langle N(\sigma) | \int d^3x \left( T^\mu_\mu(\text{QCD}) - T^\mu_\mu(D) \right) | N(\sigma) \rangle.
\label{eq:feynman_hellmann}
\end{align} 
At the vacuum $\sigma=0$, the right hand side is evaluated as a subtraction form, 
\begin{align} 
  \langle N | \int d^3x \left( T^\mu_\mu(\text{QCD}) - T^\mu_\mu(D) \right) | N \rangle \Bigg|_{\sigma=0}  
\,.   
\end{align} 
Here, the individual nucleon matrix element is dominated by the gluon condensate operator~\cite{Bali:2016lvx, FlavourLatticeAveragingGroupFLAG:2024oxs} (see also the form of $T_\mu^\mu$ in Eqs.(\ref{eq:trace_anomaly}) and (\ref{eq:trace_anomaly:mirror}), which is, in particular, in a form of what is called a disconnected graph with the factorized gluon condensate {\it irrespective to} the valence quarks which form a naive nucleon state $\sim uud, udd$~\cite{QCDSF:2012mkm, QCDSF:2020fio, QCDSF:2020den}.

Since in this symmetric limit, two QCD sectors are simply identical copies. The infrared dynamics and the gauge coupling running can therefore be fully controlled by the gluon ``decoupling'' together with the lightest quarks (e.g., see \cite{Gao:2021wun, Lu:2023mkn, Deur:2025rjo}). 
Consequently, the two sectors give the same strong coupling scale, i.e., infrared intrinsic scale $\Lambda_{\rm QCD} = \Lambda_D$. 
Hence the nonperturbative gluon loop contributions in the disconnected diagram, relevant to $m_G$, are also the same, i.e., 
\begin{align}
    \langle N | T^\mu_\mu(\text{QCD})| N \rangle \Bigg|_{\sigma=0} 
    =   \langle N |  T^\mu_\mu(D)| N \rangle \Bigg|_{\sigma=0} = m_G 
\,. 
\end{align} 
Thus, in the presence of the mirror sector along with the bi-conformal construction, the gluonic contribution to the nucleon mass $m_G$, which has been currently estimated on lattice QCD~\cite{Bali:2016lvx, FlavourLatticeAveragingGroupFLAG:2024oxs}, is interpreted as the mass identically shared by QCD and dark QCD. Note also that 
this sharing does not require an ordinary SM nucleon itself to be $Z_2$-invariant or to contain mirror constituents. The Feynman-Hellmann theorem makes this relative scalar response explicit: 
consequently, we have  
\begin{align}
    \left. \frac{\partial m_N(\sigma)}{\partial (\sigma/\sqrt{2}f_\Sigma)} \right|_{\sigma=0} = 0.
\label{eq:deriv_zero}
\end{align} 
Thus, the linear coupling of $\sigma$ to nucleons, which is proportional to this difference, vanishes identically. 
This implies the expansion around the vacuum $\langle \sigma \rangle =0$ with $\sigma = 0 + \tilde{\sigma}$,   
$m_N(\tilde{\sigma}) = m_N(0) + \frac{1}{2} m_N''(0) \tilde{\sigma}^2 + \cdots$. 
Thus the tree-level scalaron exchange amplitude for ordinary matter is exactly zero. 
This is the origin of the fifth-force cancellation in the $Z_2$ symmetric limit.

Actually, the stationary condition in Eq.(\ref{eq:deriv_zero}) should trivially be satisfied because the full theory is now invariant under $\sigma \leftrightarrow -\sigma$. This is supported by this $\sigma$-parity invariant potential form of $V_{\rm align} \sim \cosh \sigma$ in Eq.(\ref{eq:valign}) in the $Z_2$ symmetric limit with $\Lambda_{\rm QCD}=\Lambda_D$. 
Alternatively, but with a simple-minded spirit, we could make the $Z_2$-evenness of the nucleon mass manifest. 
To this end, it is useful to reparameterize the confinement scales in the chiral limit as
\begin{equation}
\Lambda_{\rm QCD} = \Lambda_* e^{s}, \qquad \Lambda_D = \Lambda_* e^{-s},
\label{eq:Lambda_parameterization}
\end{equation}
where $\Lambda_* = \sqrt{\Lambda_{\rm QCD}\Lambda_D}$ is the geometric mean which corresponds to the total vacuum energy scale arising from the alignment potential in Eq.(\ref{eq:valign}), and $s = (\sigma - \langle \sigma \rangle)/\sqrt{2}f_\Sigma$ which parametrizes the $Z_2$-odd direction. 
The $Z_2$ symmetry acts as $s \to -s$, while $\Lambda_*$ is invariant. 
Thus, the vacuum alignment condition leaves the $s$ direction flat at the classical level. The bi-conformal matching ensures that the gluonic contribution to the nucleon mass is a function of the physical confinement scale $\Lambda_*$ only:
\begin{equation}
m_G = f(\Lambda_*),
\end{equation} 
where $f$ is the same function for both the QCD and mirror QCD sectors because they are identical copies. 
Since $\Lambda_*$ is independent of $s$, we have
\begin{equation} 
\frac{\partial m_G}{\partial s} = 0.
\end{equation}
Thus, the nucleon mass $m_N(\sigma)$ is even in $\sigma$ (or equivalently, in $s$) in the $Z_2$-symmetric limit.

The standard description of nucleon properties by SM QCD remains intact; only the \textit{response} to the QCD vacuum energy is modified by the mirror sector through the alignment mechanism; hence, no modification of hadron wavefunctions or deviation in hadron experiments occurs. 
When we turn to the realistic case with $Z_2$ breaking with quark masses turned on. In that case, the cancellation, i.e., Eq.(\ref{eq:deriv_zero}) no longer works, hence a linear force coupling is recovered. As it will turn out below, the residual linear force coupling is, however, highly suppressed by the smallness of the $Z_2$ breaking, compared to the conventional dilatonic coupling to nucleon, $g_{\sigma NN}|_{\rm conventional} \sim m_N/M_P$~\cite{Fujii:2003pa, Fujii:2015iuz,Kaplan:2000hh}.

\subsection{Realistic $Z_2$ broken case} 

We now incorporate the effect of $Z_2$ breaking from quark masses. In the SM, $u$, $d$, and $s$ quarks acquire masses through the electroweak Higgs mechanism ($m_u \sim 2.5$ MeV, $m_d \sim 5$ MeV, $m_s \sim 95$ MeV). 
The mirror quarks are taken to be sufficiently light ($m_{q_D} \ll \Lambda_{\text{QCD}}$), so that their sigma terms (such as mirror counter parts of $\sigma_{\pi N}$ and $\sigma_s$) are negligible. The resulting mirror pions do not affect the fifth-force prediction; a consistent cosmological scenario exists where they are not overproduced (see also the Appendix~\ref{sec:mirror_pions}). 

First, recall the pure QCD case (without a mirror sector), where the nucleon mass decomposes as:
\begin{align}
    m_N = \langle N| T^\mu_\mu | N \rangle_{\text{QCD}} = m_G + \sigma_{\pi N} + \sigma_s,
\label{eq:mN_decomp}
\end{align}
where $m_G = \langle N| \frac{\beta(g)}{2g} G^2 | N \rangle$ is the gluonic contribution, and the sigma terms are defined as $\sigma_q = m_q \langle N| \bar{q}q | N \rangle$. Recent lattice determinations give $\sigma_{\pi N} = 35-45$ MeV and $\sigma_s = 30-50$ MeV \cite{Bali:2016lvx, FlavourLatticeAveragingGroupFLAG:2024oxs}.

When SM quark masses are introduced, the $Z_2$ breaking shifts the SM sector's matrix element by the sigma terms, while the mirror sector receives only negligible corrections (of order $m_{q_D}^2/m_G^2$). Thus, to leading order in $Z_2$ breaking, we have 
\begin{align}
    \langle N | T^\mu_\mu(\text{QCD}) - T^\mu_\mu(D) | N \rangle = \sigma_{\pi N} + \sigma_s.
\label{eq:diff_sigma}
\end{align}

From Eq.~\eqref{eq:anomaly_coupling_main}, the effective Yukawa coupling of $\sigma$ to nucleons is then evaluated as  
\begin{align}
g_{\sigma NN} &= \frac{1}{\sqrt{2}f_\Sigma} \left( \langle N | T^{\mu}_{\mu}(\text{QCD}) | N \rangle - \langle N | T^{\mu}_{\mu}(D) | N \rangle \right)  
= \frac{\sigma_{\pi N} + \sigma_s}{\sqrt{2}f_\Sigma}. 
\label{eq:g_sigma_NN_final}
\end{align} 
A linear force is thus recovered in the realistic $Z_2$ broken case. 
However, the coupling strength is dramatically suppressed by a factor of $((\sigma_{\pi N}+\sigma_s)/\sqrt{12 + 2/\xi} m_N)$, compared to the conventional dilatonic case with $g_{\sigma NN} |_{\rm conventional}\sim m_N/M_P$~\cite{Fujii:2003pa, Fujii:2015iuz, Kaplan:2000hh}. 

The residual linear force coupling in Eq.(\ref{eq:g_sigma_NN_final}) generates 
the fifth force via the scalaron $\sigma$ exchange, which yields  
the Yukawa potential $V(r) = -G_N m_N^2 (\alpha/r) e^{-m_\sigma r}$. 
The fifth-force strength $\alpha$ is then given as 
\begin{align}
\alpha = \frac{g_{\sigma NN}^2}{4\pi G_N m_N^2} 
= \frac{(\sigma_{\pi N} + \sigma_s)^2}{m_N^2} \cdot \frac{1}{6+1/\xi}, 
\label{eq:alpha_final} 
\end{align}
where we have used  $G_N = 1/(8\pi M_P^2)$ and $f_\Sigma = \sqrt{6+1/\xi}\,M_P$.  
With $\sigma_{\pi N} + \sigma_s = 65-95$ MeV from lattice determinations \cite{Bali:2016lvx, FlavourLatticeAveragingGroupFLAG:2024oxs} and $m_N \approx 940$ MeV, we have 
\begin{align}
\alpha \approx (4.8-10.2)\times 10^{-3} \times \left( \frac{1}{6+1/\xi} \right).
\label{alpha-estimate}
\end{align}
This fifth force is highly suppressed by a factor of $((\sigma_{\pi N} + \sigma_s)/\sqrt{6 + 1/\xi} m_N)^2 = {\cal O}(10^{-2})^2$ for $\xi={\cal O}(1)$, compared to the conventional dilatonic case~\cite{Fujii:2003pa, Fujii:2015iuz,Kaplan:2000hh}. 
This is the consequence of the small $Z_2$ breaking, where the ideal symmetric limit realizes the complete screening of the linear force, as has been noted in Sec.~\ref{sec:vac}. 

A key prediction of the framework is a parameter-independent correlation between the scalaron mass and the fifth-force strength. From Eqs.~\eqref{eq:general_scaling_main} and \eqref{eq:alpha_final}, eliminating $f_\Sigma$ yields 
\begin{align}
\alpha = \frac{(\sigma_{\pi N} + \sigma_s)^2}{m_N^2} \cdot \frac{(16\pi^2)^2}{\mathcal{C}} \frac{M_P^2}{{\cal M}^4} \, m_\sigma^2. 
\label{eq:correlation}
\end{align}
This correlation is independent of the non-minimal coupling $\xi$ and provides a robust, testable connection between fifth-force experiments and scalaron searches.

In the SM benchmark with $M_{\text{heavy}} = m_t \approx 173$ GeV and $\mathcal{C} \approx 196$, as in Eq.(\ref{eq:m_sigma_SM}),  
we have  
\begin{align}
\alpha|_{\rm SM} \approx (4.8 - 10.2)\times 10^{-3} \times \left( \frac{m_\sigma|_{\rm SM}}{2.7 \times 10^{-6}\text{ eV}} \right)^2, 
\label{eq:correlation_numerical}
\end{align} 
where we have used 
$\sigma_{\pi N} + \sigma_s = 65-95$ MeV~\cite{Bali:2016lvx, FlavourLatticeAveragingGroupFLAG:2024oxs}. 

The coupling $g_{\sigma NN}$ in Eq.~(\ref{eq:alpha_final}) is defined for the physical fluctuation $\tilde{\sigma} = \sigma - \langle\sigma\rangle$ around the true vacuum. Since the VEV shift $\langle\sigma\rangle/f_\Sigma \sim {\cal O}((\sigma_{\pi N} + \sigma_s)/m_N) \sim 0.09$ is small, the difference between evaluating the coupling at $\sigma=0$ and at $\sigma=\langle\sigma\rangle$ is of order $(\langle\sigma\rangle/f_\Sigma)^2 \sim 10^{-2}$, which is negligible for our parametric estimate.

\section{Prediction and Experimental Landscape}
\begin{figure}[t]
    \includegraphics[width=0.8\textwidth]{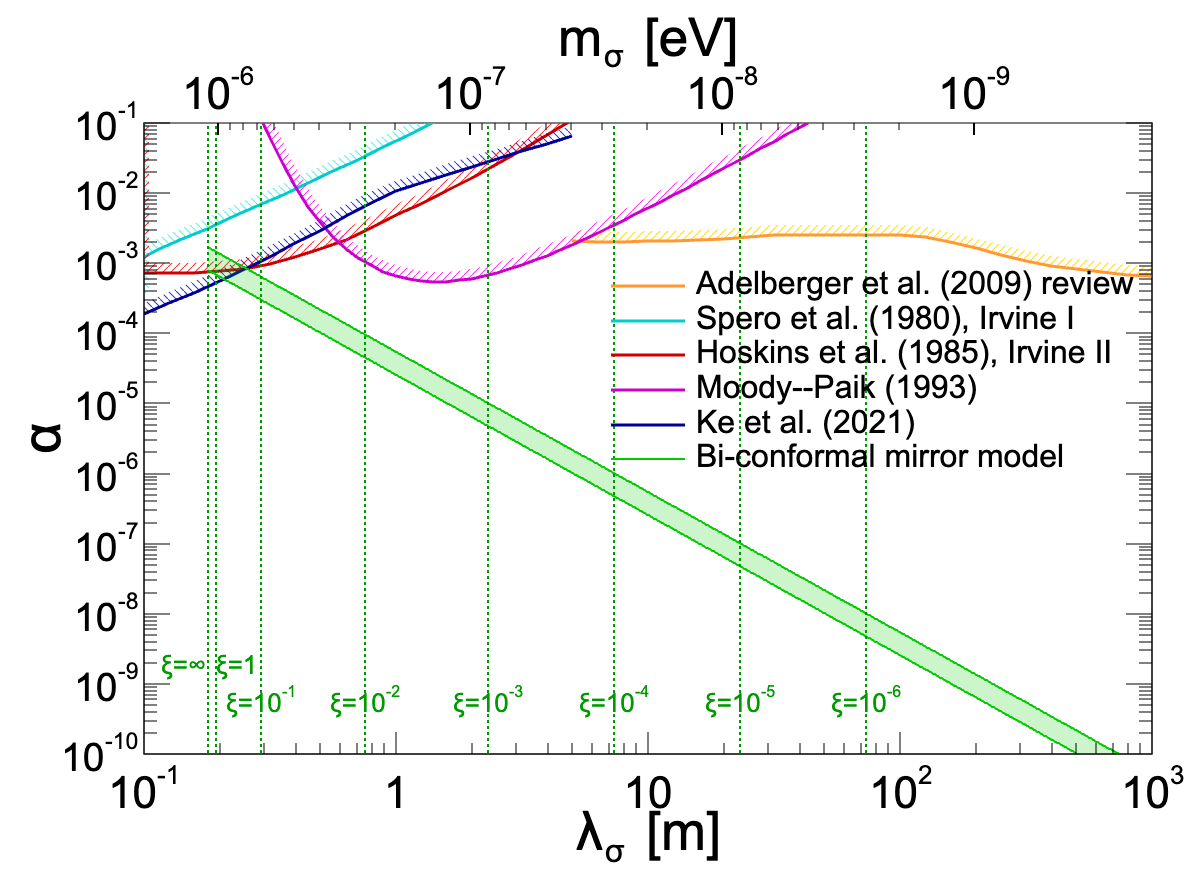}
    \caption{Prediction band of the bi-conformal mirror model in the $\alpha$--$\lambda_{\sigma}$ plane for given Compton wavelength $\lambda_{\sigma}\equiv 1/m_{\sigma}$ compared with the current upper limits on $\alpha$. The centimeter-to-meter constraints are provided by torsion-balance (Spero et al.~\cite{Spero1980} and Hoskins et al.~\cite{Hoskins1985}), gravity-gradiometer (Moody--Paik~\cite{MoodyPaik1993}), and combined $G$-measurement analyses (Ke et al.~\cite{Ke2021}). For $\lambda_{\sigma}\gtrsim 6\,\mathrm{m}$, the geophysical envelope is digitized from the review paper (Adelberger et al.~\cite{Adelberger:2009zz}), following Refs.~\cite{Adelberger2003,FischbachTalmadge1999}, and is based on lake/reservoir \cite{Moore1988,Mueller1989,Mueller1990,Cornaz1994,Hubler1995}, tower \cite{Thomas1989,Speake1990,Romaides1994,Romaides1997}, and ocean~\cite{Zumberge1991} measurements. The left edge of the band corresponds to the physical limit at $\xi=\infty$ in Eq.(\ref{eq:m_sigma_SM}) and the band width corresponds to uncertainty from lattice determinations in Eq.(\ref{alpha-estimate}).}
 \label{Fig1}
\end{figure}

Figure~\ref{Fig1} shows the prediction band of our model in the $\alpha-\lambda_{\sigma}$ plane for given Compton wavelength $\lambda_\sigma \equiv 1/m_\sigma$, where representative current laboratory and geophysical limits on $\alpha$ are displayed. The centimeter-to-meter constraints include the 2--5 cm mass-tube torsion-balance null test of Spero et al.~\cite{Spero1980}, the 5--105 cm near--far mass torsion-balance comparison of Hoskins et al.~\cite{Hoskins1985}, the meter-scale superconducting gravity-gradiometer Gauss-law test of Moody and Paik~\cite{MoodyPaik1993}, and the combined reanalysis by Ke et al.~\cite{Ke2021} of two independent torsion-pendulum measurements of $G$, which constrains Yukawa-type deviations over the 5--500 mm range. The geophysical constraint above $\lambda_{\sigma} \sim 6$~m is digitized from Adelberger et al.~\cite{Adelberger:2009zz}, whose compilation follows Adelberger, Heckel, and Nelson~\cite{Adelberger2003} and the review of Fischbach and Talmadge~\cite{FischbachTalmadge1999}. The underlying terrestrial measurements include lake/reservoir experiments~\cite{Moore1988,Mueller1989,Mueller1990,Cornaz1994,Hubler1995}, tower experiments~\cite{Thomas1989,Speake1990,Romaides1994,Romaides1997}, and an ocean experiment~\cite{Zumberge1991}. 
The scalaron mass $m_\sigma$ predicted from the SM, Eq.(\ref{eq:m_sigma_SM}), is itself sensitive to the nonminimal coupling $\xi$.  The left edge of the band corresponds to the physical limit at $\xi=\infty$ in Eq.(\ref{eq:m_sigma_SM}). Current cosmological constraints on the non-minimal coupling $\xi$ are model-dependent. For conventional scalar-tensor inflation models, Planck 2018 places an upper bound $\xi < 0.089$ at 95\% CL~\cite{Planck:2018vyg}. In our scenario, $\sigma_+$ is assumed to be stabilized at its $Z_2$-symmetric VEV with Planck-scale mass, and its inflationary dynamics are not specified. If this Planck limit is simply applied to Eq~\eqref{eq:m_sigma_SM}, 
we could have 
$m_\sigma < 6.5 \times 10^{-7}$ eV, and the $\xi$-free $\alpha- m_\sigma$ correlation in Eq.~\eqref{eq:correlation_numerical} gives $\alpha < (2.8 - 6.0) \times 10^{-4}$. 

The cosmological evolution of $\sigma$ is also consistent with the bound from 
the electron-to-proton mass ratio $\mu \equiv m_e/m_p$.  
As shown in the Appendix \ref{sec:mu_stability}, the cosmologically consistent scenario yields $\dot{\sigma}\ll H_0 M_P$ in the present day, giving $\dot{\mu}/\mu \sim 10^{-15} \, {\rm yr}^{-1}\times (\dot{\sigma}/H_0 M_P) \sim 0$, which is thus well within current experimental bounds \cite{Rosenband:2008qgq}.  
The $\mu$-variation bound should be viewed as an important time-variation consistency check on the aligned vacuum structure rather than as a substitute for a dedicated analysis of static equivalence-principle observables. In the present framework, 
both the leading $\sigma$-dependence of $m_p$ and the dominant residual nucleon coupling are controlled by the approximately universal combination $\sigma_{\pi N} + \sigma_s$. 
 
Future atom-interferometric facilities such as AION~\cite{Badurina:2019hst} and MAGIS-100~\cite{Abe:2021magis100} provide strong motivation for revisiting this region experimentally, although their published sensitivity projections are typically presented either for dark-matter-induced signals or for composition-dependent vector interactions rather than directly as bounds on the universal Yukawa parameter $\alpha$ adopted here. We therefore regard these facilities as promising platforms for future meter-scale searches, without claiming a one-to-one published projection in the $(\alpha, \lambda)$ plane for the specific universal scalar force considered in this work. 

The scalaron couples to photons through the electromagnetic trace anomaly term as (see also Eq.(\ref{eq:trace_anomaly}) in the Appendix~\ref{sec:geometric_derivation}) 
\begin{align}
\mathcal{L}_{\sigma FF} = \frac{\sigma}{\sqrt{2}f_\Sigma} \left[ -\frac{\beta(e)}{2e} F_{\mu\nu}F^{\mu\nu} \right] 
\equiv 
g_{\sigma \gamma\gamma}  \left[ -\frac{1}{4} \sigma F_{\mu\nu}F^{\mu\nu} \right], 
\label{sigma-photon}
\end{align} 
where $F_{\mu\nu}$ stands for the field strength of the photon field and 
$\beta(e) = 11 e^3/(32\pi^2)$ is the beta function of the electromagnetic coupling $e$, which includes the full one-loop SM corrections. 
This yields the effective coupling $g_{\sigma \gamma\gamma}$, 
\begin{align}
g_{\sigma\gamma\gamma} = \frac{11\alpha_{\text{EM}}}{4\pi\sqrt{2}f_\Sigma} 
\approx 1.9 \times 10^{-21}\,\text{GeV}^{-1} \times \left(\frac{1}{\sqrt{6+1/\xi}}\right). 
\label{eq:g_sigma_gamma}
\end{align}
This can further be rewritten as a function of the SM prediction for $m_\sigma$ given in Eq.(\ref{eq:m_sigma_SM}), freely from the parameter $\xi$: 
\begin{align}
g_{\sigma\gamma\gamma}  
\approx 1.9 \times 10^{-21}\,\text{GeV}^{-1} \times \left( \frac{m_\sigma|_{\rm SM}}{2.7 \times 10^{-6} \, {\rm eV}} \right). 
\label{eq:g_sigma_gamma-m_sigma}
\end{align}

The Stimulated Resonant Photon Collision (SRPC) technique \cite{Homma:2019rqb, Katsuragawa:2021wmw} provides a direct probe of this coupling through resonant photon-photon scattering when the center-of-mass energy matches $m_\sigma$. The sensitivity of SRPC can reach $g_{\sigma\gamma\gamma} \sim 10^{-22}-10^{-21}$ GeV$^{-1}$ at $m_\sigma \sim 10^{-7}-10^{-6}$ eV, which enables tests at $\xi=\infty-10^{-2}$ as shown in Fig.~\ref{Fig1}.

\section{Conclusion and Outlook}

We have presented a novel, technically natural screening mechanism for fifth forces mediated by light scalars. The proposed mechanism demonstrates that fundamental principles of particle physics, such as symmetry, anomaly structure, and vacuum alignment, can provide natural solutions to the fifth-force problem without resorting to environmental screening or fine-tuning. This establishes a new ballpark for connecting particle physics to gravitational phenomenology, with testable predictions that will be probed by upcoming experiments across multiple frontiers. The scaling formula $\alpha \times (m_\sigma {\cal M}/M^2_P)^2 \sim \text{constant}$ predicts $m_\sigma \lesssim 10^{-6}$ eV and $\alpha \lesssim 10^{-3}$ for the SM. This places the prediction directly in the target range of next-generation fifth-force and stimulated photon-photon collider experiments. 

The present work also opens up several new directions. The screening mechanism applies to any light scalar with Planck-suppressed couplings to matter. It provides a natural solution to the fifth-force problem for light dark-matter candidates (axion-like particles) and for moduli fields in string-theory compactifications. The mechanism can be generalized to other gauge groups and matter content. 

While the proposed model offers a natural solution to the fifth-force problem, the ultimate origin of the Planck-scale VEVs and the full cosmological history of the scalaron are subjects for the future work. Crucially, these UV and cosmological details do not affect the robust, testable low-energy prediction for fifth-force strength.

\section*{Acknowledgments}

This work is dedicated to the late Professor Yasunori Fujii, whose pioneering works on dynamical dark energy models have inspired this line of research. 
K.H. acknowledges the support of the Collaborative Research Program of the Institute for Chemical Research at Kyoto University (Grant No. 2025-100), the JSPS Core-to-Core Program (grant number: JPJSCCA20230003), and Grants-in-Aid for Scientific Research (Nos. 24KK0068 and 25H00645) from the Ministry of Education, Culture, Sports, Science and Technology (MEXT) of Japan.
T.K. is supported by the National Natural Science Foundation of China (No.~12403003) and National Key R\&D Program of China (No.~2021YFA0718500).

\appendix

\section{Derivation of the Anomaly Coupling}
\label{sec:geometric_derivation}

We provide a derivation of the scalaron coupling to the difference of QCD nonperturbative trace anomalies using the geometric properties of the bi-conformal construction and the exact quantum definition of the energy-momentum tensor. 

We begin with the Jordan-frame actions for the relevant parts in SM and mirror sectors:
\begin{align}
\mathcal{L}_J^{\text{SM}} &= \sqrt{-g}\left[\frac{1}{2}\xi\phi^2 R(g) + \frac{1}{2}(\partial\phi)^2 + \mathcal{L}_{\text{SM matter}}(g,\psi)\right], \notag \\
\mathcal{L}_J^{\text{Mirror}} &= \sqrt{-g_D}\left[\frac{1}{2}\xi\phi_D^2 R_D(g_D) + \frac{1}{2}(\partial\phi_D)^2 + \mathcal{L}_{\text{Mirror}}(g_D,\psi_D)\right]. \label{eq:J_mirror}
\end{align}
The two Jordan-frame metrics are related by the bi-conformal constraint $\Omega^2 g_{\mu\nu} = \Omega_D^2 g_{D\mu\nu}$, where the Weyl factors are defined as 
\begin{align}
\Omega^2 = \frac{\xi\phi^2}{M_P^2}, \qquad \Omega_D^2 = \frac{\xi\phi_D^2}{M_P^2}.
\end{align}
We transform both sectors to a common Einstein-frame metric $g_{\mu\nu}^E$ via:
\begin{align}
g_{\mu\nu}^E = \Omega^2 g_{\mu\nu} = \Omega_D^2 g_{D\mu\nu}.
\end{align}

For any matter action $S_{\text{matter}}[g,\psi]$ coupled to a metric $g_{\mu\nu}$, under a Weyl transformation $g_{\mu\nu} \to g^{E}_{\mu\nu} = e^{2\alpha(x)}g_{\mu\nu}$, the matter action transforms accordingly.
To linear order around the background satisfying $\Omega=1$, we obtain a relation between the matter actions in two frames:
\begin{align}
S_{\text{matter}}[g^{E}_{\mu\nu},\psi] 
= 
S_{\text{matter}}[g_{\mu\nu},\psi] 
+ \int d^4x \sqrt{-g} \, \alpha(x) \, T^\mu_\mu(x) + \mathcal{O}(\alpha^2),
\label{eq:weyl_variation}
\end{align}
where $T^{\mu\nu}(x) = \left(2/\sqrt{-g} \right) \left( \delta S_{\text{matter}}/\delta g_{\mu\nu}(x)\right)$ is the exact quantum energy-momentum tensor operator. 
Here, $T_{\mu\nu}$ conventionally denotes the energy-momentum tensor in the Jordan frame because it is defined by the Jordan frame metric $g_{\mu\nu}$.
To linear order around the background $\Omega=1$, however, the distinction between the Jordan- and Einstein-frame energy-momentum tensors affects only higher-order terms in $\alpha(x)$.
Hereafter, we work entirely with the Einstein-frame matter fields and denote the corresponding energy-momentum tensor simply by $T_{\mu\nu}$.

From the definition of the Weyl factor, we identify $\alpha(x)$ as:
\begin{align}
\alpha(x) = \frac{1}{2}\ln\left(\frac{\xi\phi^2(x)}{M_P^2}\right).
\label{eq:alpha_exact}
\end{align}
Expanding around the vacuum expectation value (VEV) $\langle\phi\rangle = v_\phi = M_P/\sqrt{\xi}$ with $\phi = v_\phi + \delta\phi$:
\begin{align}
\alpha(x) = \frac{\delta\phi}{v_\phi} - \frac{1}{2}\frac{\delta\phi^2}{v_\phi^2} + \cdots
\label{eq:alpha_expansion}
\end{align}

The fluctuation $\delta\phi$ is not canonically normalized. After Weyl transformation to the Einstein frame, the canonically normalized field $\Sigma$ is related to $\delta\phi$ by:
\begin{align}
\Sigma = \sqrt{6 + \frac{1}{\xi}} \, M_P \, \frac{\delta\phi}{v_\phi} + \mathcal{O}(\delta\phi^2).
\label{eq:canonical_normalization}
\end{align}
We define the physical coupling scale:
\begin{align}
f_\Sigma \equiv \sqrt{6 + \frac{1}{\xi}} \, M_P, 
\label{fsigma}
\end{align}
so that $\delta\phi/v_\phi = \Sigma/f_\Sigma + \mathcal{O}(\Sigma^2)$. Substituting into Eq.~\eqref{eq:alpha_expansion}:
\begin{align}
\alpha(x) = \frac{\Sigma(x)}{f_\Sigma} + \mathcal{O}(\Sigma^2/f_\Sigma^2).
\label{eq:alpha_sigma}
\end{align} 

For the mirror sector, we similarly have $\alpha_D(x) = \Sigma_D/f_\Sigma + \cdots$, with the same $f_\Sigma$. The light scalaron field $\sigma$ is the $Z_2$-odd combination: $\Sigma = \sigma/\sqrt{2}$, $\Sigma_D = -\sigma/\sqrt{2}$. Thus, the linear coupling to the trace anomalies is:
\begin{align}
\mathcal{L}_{\text{int}} = \frac{\sigma}{\sqrt{2}f_\Sigma} \left[ T^\mu_\mu(\text{SM}) - T^\mu_\mu(\text{Mirror}) \right].
\label{eq:final_interaction}
\end{align}

The SM trace anomaly in the electroweak (EW) symmetric phase is given by the exact operator:
\begin{align}
T^\mu_\mu \Bigg|_{\rm SM}^{v_{\rm EW}=0}  = \frac{\beta(g_s)}{2g_s} G^a_{\mu\nu}G^{a\mu\nu}  
+ \frac{\beta(g_W)}{2g_W} G^i_{\mu\nu}W^{i \mu\nu} 
+\frac{\beta(g_Y)}{2g_Y} B_{\mu\nu}B^{\mu\nu} 
+ (2 - \gamma_{\Phi^2}) m^2_\Phi |\Phi|^2\,,   
\label{eq:trace_anomaly:high}
\end{align}
where $g_s, g_W, g_Y$, and $m_\phi$ are $SU(3)_c$, $SU(2)_W$, $U(1)_Y$, and the Higgs mass parameter, respectively, and $\beta(g_{s,W,Y})$ and $\gamma_{\Phi^2}$ are corresponding beta functions (anomalous dimensions). Here the sums over repeated indices $a=1, \cdots , 8$ for the adjoint representation of $SU(3)$ and $i = 1,2,3$ for the adjoint representation of $SU(2)$ have been taken into account. At scales down to low-energy $\mu = \Lambda_{\rm QCD} \sim 0.3$ GeV, where heavier particles are decoupled, in which the Higgs only leaves its VEV $v_{\rm EW} = 246$ GeV, the anomaly form is reduced to 
\begin{align}
    T^\mu_\mu \Bigg|_{\rm SM}^{\Lambda_{\rm QCD}} 
    = \frac{\beta(g_s)}{2g_s} G^a_{\mu\nu}G^{a\mu\nu}  
    +\frac{\beta(e)}{2 e} F_{\mu\nu} F^{\mu\nu} 
    + \sum_{q=u,d,s} (3 - \gamma_m^{q}) m_q \bar{q}q  
    \,, 
\label{eq:trace_anomaly} 
\end{align}
where $\gamma_m^q$ denotes the mass anomalous dimension of $m_q$, and the electromagnetic (EM) coupling $e$ arises in the standard way as in the SM: $1/e^2 = 1/g_W^2 + 1/g_Y^2$. Since the trace anomaly is renormalization group (RG) invariant, we have $T^\mu_\mu |_{\rm SM}^{v_{\rm EW}=0}  =  T^\mu_\mu |_{\rm SM}^{\Lambda_{\rm QCD}}$ in a sense of the RG evolution, where quantum corrections from decoupled particles are absorbed into the anomalous dimensions (beta functions) and the Higgs VEV contributes to the $u, d, s$ quark masses. Thus, $\beta(g_s)$ is the full QCD beta function, which is at low energy, governed by the nonperturbative QCD dynamics to determine the intrinsic scale $\Lambda_{\rm QCD}$ as the hadronic scale. 

The mirror sector yields a similar trace anomaly, but without the EM contributions: 
\begin{align}
    T^\mu_\mu \Bigg|_{\rm Mirror} 
    = \frac{\beta(g_D)}{2g_D} G^a_{D \mu\nu}G_D ^{a\mu\nu}  
    + \sum_{q_D=u_D,d_D,s_D} (3 - \gamma_{m_D}^{q_D}) m_{q_D} \bar{q_D}q_D   
    \,.  
\label{eq:trace_anomaly:mirror} 
\end{align}
Here, the mirror quark masses also come in, which can be generated by quantum gravity effect to give $m_{q_D} \ll m_q$ (See the Appendix~\ref{sec:mirror_pions}).

\section{Scalaron $\sigma$ Mass via Finite Two-Loop Mechanism}
\label{sec:mass_detail}

The mass of the scalaron $\sigma$, being a pseudo Nambu-Goldstone boson of the scale symmetry, is protected by the shift symmetry $\sigma \to \sigma + c$. This shift symmetry is anomalous and broken by the $\sigma$ couplings to the SM sector at the loop level. The symmetry violation becomes evident at the two-loop level, which turns out to be non-divergent. We derive the finite two-loop mass for $\sigma$ in the true vacuum after alignment. The calculation assumes that $\sigma$ is already stabilized at its minimum $\langle \sigma \rangle = f_\Sigma/\sqrt{2} \ln (\Lambda_{\rm QCD}/\Lambda_D)$ determined by Eq.~\eqref{eq:sigma_min} of the main text, and that all SM particles have their usual masses from EW symmetry breaking. The mass $m_\sigma$ is then the curvature of the effective potential at this minimum:
\begin{align}
m_\sigma^2 = \left. \frac{d^2 V_{\text{eff}}}{d\sigma^2} \right|_{\sigma = \langle \sigma \rangle}.
\end{align}

\subsection{Universal Scaling Relation for Scalaron Mass}

The interaction Lagrangian originates from the conformal factors $\Omega^2 = \xi\phi^2/M_P^2$ and $\Omega_D^2 = \xi\phi_D^2/M_P^2$ in the Jordan frames. Transforming to the Einstein frame and expanding around the true vacuum $\langle \phi \rangle = \langle \phi_D \rangle = M_P/\sqrt{\xi}$, the coupling to $\sigma$ in $D = 4-\epsilon$ dimensions takes the universal form:
\begin{align}
    \mathcal{L}_{\text{int}} 
    &\supset 
    (D-4)\frac{\sigma}{\sqrt{2}f_\Sigma} \, {\cal M} \, \bar{\psi}\psi 
    \quad \text{(for fermion $\psi$)},
    \notag \\
    \mathcal{L}_{\text{int}} 
    &\supset 
    (D-4)\frac{\sigma}{\sqrt{2}f_\Sigma} \left(\frac{1}{2} {\cal M}^2 V^2\right) 
    \quad \text{(for boson $V$)},
\label{eq:universal_coupling}
\end{align}
where $\mathcal{M}$ represents the characteristic mass scale of particles, 
and we have kept only the linear $\sigma_-$ coupling. The crucial $\epsilon$ factor ensures that tree-level couplings vanish in $D=4$, making the leading contribution to $m_\sigma$ arise at two-loop order. This is independent of whether $\sigma$ is rolling or static—it is a property of the Feynman rules in the quantum theory.

The scalaron mass squared $m_\sigma^2$ is given by the vacuum polarization $\Pi_{\sigma\sigma}(p^2)$ at $p^2=0$. Since each $\sigma$ vertex carries a factor $\propto \epsilon$, the leading contribution arises at two-loop order. For any diagram with two external $\sigma$ legs and two mass insertions from heavy particles, the amplitude has the schematic form:
\begin{align}
    -i \Pi_{\sigma\sigma}(0) 
    \sim 
    \frac{\epsilon^2}{2 f_\Sigma^2} 
    \times \left( \frac{A}{\epsilon^2} + \frac{B}{\epsilon} + C \right) 
    \times \mathcal{M}^4,
\end{align}
where $A, B, C$ are dimensionless constants from loop integration. The $\epsilon^2$ from the vertices exactly cancels the leading $(1/\epsilon)^2$ factor, yielding a finite result proportional to $A \mathcal{M}^4/f_\Sigma^2$ as $\epsilon \to 0$.

Thus, independent of the specific diagram topology, the scalaron mass squared scales as:
\begin{align}
m_\sigma^2 \sim \frac{1}{(16\pi^2)^2} \frac{\mathcal{M}^4}{f_\Sigma^2} \times (\text{coupling factors}) \times (\text{color/flavor factors}),
\label{eq:general_scaling}
\end{align}
where the $(16\pi^2)^{-2}$ factor is characteristic of two-loop diagrams.

\subsection{Contributions from SM Particles}

We estimate the dominant contribution to $m_\sigma$ from SM particles. A complete two-loop calculation requires careful treatment of gauge dependence and counterterms, which we leave for future work. However, the parametric scaling $m_\sigma^2 \sim m_t^4/(16\pi^2 f_\Sigma)^2$ is robust on dimensional grounds and from the structure of the anomaly-induced coupling. Establishing the $m_\sigma$-$\alpha$ correlation in Eq.~\eqref{eq:general_scaling} is the key phenomenological prediction. 

In the SM, all massive particles acquire mass through EW symmetry breaking and couple to $\sigma$ via Eq.~\eqref{eq:universal_coupling}. The calculation is performed in the true vacuum where $v_{\text{EW}} = 246$ GeV, giving the top quark, $W$, $Z$, and Higgs their observed masses. The dominant contributions come from the top quark due to its large Yukawa coupling and mass. Gauge boson contributions ($W$, $Z$) are subdominant, typically suppressed by factors of $(M_W/m_t)^4 \sim 0.05$, and smaller coupling constants. The Higgs loop contributions are also suppressed due to the small quartic coupling $\lambda_h = (m_h/2 v_{\rm EW})^2 \sim 0.012$, which leads to the suppression by a factor of $\sim (\lambda_h/y_t)^2 \sim 10^{-2}$. 

Thus, the leading contributions arise from two-loop diagrams involving the top quark with the SM Higgs sector or gluons. Among those, it turns out that the diagrams having the top quark propagators dressed by the SM Higgs and gluon arches, as illustrated in Fig.\ref{Fig2}, only give the loop integral structure yielding the factor $(1/\epsilon)^2$: 
Note that the two-loop correction to $m_\sigma$ is controlled by UV double-log divergences; hence, the loop integrals are dominated at high energies ($k \gg M_W$), and the Goldstone equivalence theorem can be applied there. 
Therefore, the gaugeless limit (Higgs ($h$) + Nambu-Goldstone bosons ($\pi^{\pm, 0}$)) correctly captures the gauge-invariant result. The factor of 3 from loops with $h$, $\pi^0$, $\pi^-$ is then constructively added up; the $\pi^+$ loop is negligible due to the small bottom Yukawa coupling. 

The relevant loop integral $I$ is evaluated as  
\begin{align} 
I =  \int \frac{d^D k}{(2\pi)^D} \int \frac{d^D p}{(2\pi)^D} \frac{k^4}{(k^2 - m_t^2)^3 ((k+p)^2 - m_t^2) (p^2- M^2)} 
 \,, 
\end{align}
where $M=m_h (0)$ for the Higgs (gluon and the Nambu-Goldstone bosons) case, and we have omitted $i\epsilon$ terms for propagators in the denominator. This $I$ forms $m_\sigma^2$ as $m_\sigma^2 = I \times$ (overall couplings) $\times$ (group factors). First, the $p$-integration simply gives a logarithmic divergence: 
\begin{align}
  \int \frac{d^D p}{(2\pi)^D} \frac{1}{ ((k+p)^2 - m_t^2) (p^2- M^2)}  \approx    \frac{i}{16\pi^2} \frac{2}{\epsilon}\,, 
\end{align} 
independent of $k$ and $M$, and then the loop integral function $I$ goes like  
\begin{align}
I \approx \frac{i}{16\pi^2} \frac{2}{\epsilon} \cdot 
  \int \frac{d^D k}{(2\pi)^D} \frac{k^4}{(k^2-m_t^2)^3} \,. 
\end{align} 
The $k$-integration yields 
\begin{align}
    I \approx  -  \frac{3m_t^2}{(16\pi^2)^2} \left( \frac{2}{\epsilon} \right)^2  \,. 
\end{align}
Combining with the overall couplings and group factors, and taking into account two ways of dressing the top propagators by the Higgs (along with $\pi^{-, 0}$) and gluon arches, we find the top-gluon mixed two-loop diagram to yield  
\begin{align} 
m_\sigma^2|_{\text{top-gluon}}= \frac{1}{(16\pi^2)^2} \frac{64\,g_s^2\,m_t^4}{f_\Sigma^2}, 
\end{align} 
and the top-Higgs sector diagram to give  
\begin{align}
m_\sigma^2|_{\text{top-Higgs sector}} = \frac{1}{(16\pi^2)^2} \frac{3 \times 36 \, y_t^2\,m_t^4}{f_\Sigma^2}.
\end{align}  
Combing two terms, we see the factor ${\cal C}$ defined in Eq.~\eqref{eq:general_scaling_main} of the main text to take ${\cal C} = 64 g_s^2 + 108 y_t^2$. 
The result can be summarized by a compact formula that takes the form as given in Eq.~\eqref{eq:m_sigma_SM} of the main text,  
\begin{align}
m_\sigma 
\approx \frac{2.7\times 10^{-6} \text{ eV}}{\sqrt{6+1/\xi}},  
\label{eq:m_sigma_compact}
\end{align} 
where we have used $f_\Sigma = \sqrt{6+1/\xi}~M_P$ and taking $v_{\rm EW} =246$ GeV, $M_P =  2.4\times 10^{18}$ GeV, $m_t = y_t v_{\rm EW}/\sqrt{2} = 173$ GeV, and $\alpha_s = g_s^2/(4\pi) = 0.11$, from which we also get ${\cal C}\approx 196$.  

\begin{figure}[t]
    \includegraphics[width=0.3\textwidth]{Fig2.png}
    \caption{The relevant two-loop graphs having the top quark propagators dressed by the SM Higgs (along with the would-be Nambu-Goldstone bosons) and gluon arches, in which the loop-integral structure yields the $(1/\epsilon)^2$ contribution. The external momentum for $\sigma$ has been set to 0.}
\label{Fig2}
\end{figure}

The scale anomaly couplings to the gluon field strengths and quark mass terms in Eq.~\eqref{eq:trace_anomaly} could, in principle, contribute to the $\sigma$ mass generation. The gluonic contribution precisely yields the terms involving the confinement scales $\Lambda_{\text{QCD}}$ and $\Lambda_D$: $m_\sigma^2|_{\text{QCD+Mirror}} \sim \Lambda_{\text{QCD}}^4/f_\Sigma^2$. The quark mass terms, by contrast, are subtracted as perturbative divergent contributions in defining the physical trace anomaly after QCD confinement, as noted in the main text; they therefore do not contribute to the curvature of the low-energy potential.

The QCD trace anomaly terms thus play the crucial role in establishing the true vacuum of $\sigma$ via vacuum alignment, while the dominant contribution to $m_\sigma$ comes from higher-mass particles such as the top quark and the SM Higgs, as discussed above.

\subsection{Renormalization Group Invariance and Threshold Matching}

The scalaron mass receives contributions from physics at widely separated scales: the QCD intrinsic scale, $\Lambda_{\text{QCD}} \sim 0.3$ GeV, and the EW physics scale, $\sim 100$ GeV. A consistent, effective field theory treatment requires that physical quantities be independent of the RG scale $\mu$, with threshold corrections properly matching the theory above and below each mass scale.

Below the EW scale, heavy particles (such as the top quark, $W$, $Z$, and the Higgs) are integrated out at their respective mass thresholds. The matching condition at $\mu = m_t$ (and similarly at $M_W$, $M_Z$, $m_h$) ensures that the effective potential is continuous across thresholds. This yields the infrared (IR) effective potential:
\begin{align}
V_{\text{eff}}^{\text{(IR)}}(\sigma) 
    = 
    \frac{\kappa}{4} \left( \Lambda_{\text{QCD}}^4 e^{-2\sqrt{2}\sigma/f_\Sigma} + \Lambda_D^4 e^{+ 2\sqrt{2}\sigma/f_\Sigma} \right) 
    + \frac{1}{2} \left[ \frac{64 g_s^2+108 y_t^2}{(16\pi^2)^2} \frac{m_t^4}{f_\Sigma^2} + \cdots \right] \sigma^2 
    + \cdots.
\label{eq:matching_condition}
\end{align}
The first and second terms come from the QCD and mirror QCD trace anomaly couplings in Eqs.(\ref{eq:trace_anomaly}) and (\ref{eq:trace_anomaly:mirror}) with the IR gluon and chiral condensates, $\langle G^2 \rangle$ and $\langle \bar{q}q \rangle$. The mass term contributions arise in the same form as the universal scaling, in Eq.(\ref{eq:general_scaling}), as noted below Eq.(\ref{eq:m_sigma_compact}), which gives positive mass squared terms. Hence, the overall coefficient $\kappa$ in Eq.(\ref{eq:matching_condition}) is positive, and the size of $\kappa$ is to be determined by non-perturbative QCD dynamics with a dilatonic background. When we simply refer to the vacuum energy scale of QCD $\sim (1 \,{\rm GeV})^4$ together with $\Lambda_{\rm QCD} \sim 0.3$ GeV, the size of $\kappa$ is expected to be of ${\cal O}(10)$. Note also that this term contribution is RG-invariant by construction, as $\Lambda_{\text{QCD}}$ is defined as the RG-invariant scale via the RG equation.

The second term comes from integrating out top quarks at two-loop order, with the specific combination $(64\alpha_s+108 y_t^2)$ reflecting the RG-invariant combination of couplings {\it frozen} at the matching scale, $\mu =m_t$. The ellipsis denotes subleading contributions from $W$, $Z$, and Higgs loops, which are numerically smaller.

The QCD and mirror QCD terms are actually negligible compared to the EW threshold corrections, which are dominated by the top loop contribution. Numerically, $\Lambda_{\text{QCD}}^4/f_\Sigma^2 \sim (10^{-10} \text{ eV})^2$, many orders of magnitude below the $10^{-6}$ eV scale from top loops. Note that this EW-origin mass of $\sigma$ is defined at the true vacuum and is not realized unless $\sigma$ reaches there.

\subsection{Vacuum Structure and the Two-Loop Calculation}

A potential conceptual concern arises from the sequencing of scales: the two-loop calculation of $m_\sigma$ from SM particles assumes the EW symmetry is broken with $v_{\text{EW}} = 246$ GeV, giving the top quark, $W$, $Z$, and Higgs their observed masses. When the QCD confinement takes place to dynamically generate the scale of $\Lambda_{\rm QCD}$ and define the vacuum, those perturbative corrections of the scale violation to the potential of $\sigma$ are subtracted, as noted in the main text. This subtraction procedure determines the standard of the potential to match the observed energy density coupled to $\sigma$, that is, the vacuum, when the fifth-force coupling is measured. Only after $\sigma$ has settled at $\langle \sigma \rangle = \frac{f_\Sigma}{\sqrt{2}} \ln (\Lambda_{\rm QCD}/\Lambda_D)$ (see Eq.(\ref{eq:sigma_min}) of the main text) do its quantum fluctuations probe the virtual effects of those massive SM particles. These virtual contributions generate the curvature of the potential at the minimum, namely the physical mass $m_\sigma$, regardless of when those particles became massive in cosmic history.

The situation is analogous to computing the mass of the Higgs boson in the SM: the one-loop top quark contribution to $m_h^2$ assumes the top is massive, but this does not require the top to have been massive during the EW phase transition. The calculation is a property of the final vacuum, not a statement about cosmological evolution.

Thus, the two-loop result in Eq.(\ref{eq:m_sigma_compact}) is valid as the curvature of the $\sigma$ potential in the present-day vacuum, independent of the thermal history that led to this vacuum. (As will be presented in the Appendix~\ref{sec:mu_stability}, the $\sigma$ field starts to roll down from $\sigma=0$ immediately after the vacuum alignment is established, and promptly reaches the true vacuum until today.) The only requirement is that the vacuum alignment minimum $\langle \sigma \rangle$ is indeed the location around which we expand—a condition satisfied by construction.

\section{Stability of the Electron-to-Proton Mass Ratio}
\label{sec:mu_stability}

A crucial constraint on dynamical scalars comes from the observed stability of the electron-to-proton mass ratio $\mu \equiv m_e/m_p$. Current atomic clock experiments constrain $|\dot{\mu}/\mu| < 1.0 \times 10^{-16}~\text{yr}^{-1}$ \cite{Rosenband:2008qgq}. 

The electron mass $m_e$ acquires $\sigma$ dependence through the electromagnetic trace anomaly. The effective $\sigma ee$ coupling arises at two-loop order via the anomaly-induced $\sigma\gamma\gamma$ vertex in Eq.(\ref{sigma-photon}) of the main text and a chirality-flipping electron loop. This yields 
\begin{align}
    g_{\sigma ee} 
    \approx
    \frac{3\alpha^2_{\rm EM}}{8\pi^2} \frac{m_e}{\sqrt{2}f_\Sigma} 
    \approx 
    2 \times 10^{-6} \times \frac{m_e}{\sqrt{2}f_\Sigma}.
\label{eq:gsigmaee}
\end{align}
Consequently,
\begin{align}
    \frac{\partial \ln m_e}{\partial (\sigma/\sqrt{2}f_\Sigma)} 
    = 
    \frac{g_{\sigma ee} \sqrt{2}f_\Sigma}{m_e} \approx 2 \times 10^{-6}.
\end{align}  
For the proton, the $\sigma$ dependence is set by the residual fifth-force coupling derived in the main text. From Eqs.~(\ref{eq:g_sigma_NN_final}) and (\ref{eq:alpha_final}), we have 
\begin{align}
    \frac{\partial \ln m_p}{\partial (\sigma/\sqrt{2}f_\Sigma)} 
    = 
    \frac{g_{\sigma pp} \sqrt{2}f_\Sigma}{m_p} 
    = 
    \frac{\sigma_{\pi N} + \sigma_s}{m_p} \approx 0.09,
\label{eq:dlnmp_dsigma}
\end{align}
where the numerical value uses the central lattice QCD estimate $(\sigma_{\pi N} + \sigma_s)/m_p \approx 85\ \text{MeV}/940\ \text{MeV} \approx 0.09$~\cite{Bali:2016lvx, FlavourLatticeAveragingGroupFLAG:2024oxs}.  
The sensitivity of the electron-to-proton mass ratio $\mu \equiv m_e/m_p$ is therefore:
\begin{align}
    \frac{\partial \ln \mu}{\partial (\sigma/\sqrt{2}f_\Sigma)} 
    = \frac{\partial \ln m_e}{\partial (\sigma/\sqrt{2}f_\Sigma)} 
    - \frac{\partial \ln m_p}{\partial (\sigma/\sqrt{2}f_\Sigma)} 
    \approx 2 \times 10^{-6} - 0.09 
    \approx -0.09,
\label{eq:dlnmu_dsigma}
\end{align}
where the proton contribution dominates by four orders of magnitude. Thus, the resulting variation is:
\begin{align}
    \frac{\dot{\mu}}{\mu} 
    = \frac{\partial \ln \mu}{\partial (\sigma/\sqrt{2}f_\Sigma)} \cdot \frac{\dot{\sigma}}{\sqrt{2}f_\Sigma}
    \approx 
    - 0.09 \times \frac{\dot{\sigma}}{\sqrt{2}f_\Sigma}.
\label{eq:mu_dot_final}
\end{align} 

Now, we consider the cosmological evolution of $\sigma$. Before the QCD phase transition ($T \gtrsim 200$ MeV), the condensates vanish and $V_{\text{align}}=0$; $\sigma$ is massless. We assume that the field is initially at $\sigma=0$, an extremum of the $Z_2$-symmetric potential. At confinement, $V_{\text{align}}$ develops a minimum at $\langle \sigma \rangle = (f_\Sigma/\sqrt{2})\ln(\Lambda_D/\Lambda_{\text{QCD}})$, with $\langle \sigma \rangle/f_\Sigma \sim (\sigma_{\pi N}+\sigma_s)/m_N \sim 0.05-0.10$ (see Sec.~\ref{sec:5th-sup}). The field rolls to this minimum on a timescale much shorter than the Hubble time. Upon reaching the minimum, the two-loop mechanism (Sec.~\ref{sec:mass_detail}) generates $m_\sigma \sim 10^{-6}$ eV (for $\xi \sim \mathcal{O}(1)$). Since $m_\sigma \gg H$ at $T \sim \Lambda_{\text{QCD}}$, oscillations are rapidly damped by Hubble friction. By Big Bang Nucleosynthesis, $\sigma$ is static at $\langle \sigma \rangle$, ensuring that late-time variations of fundamental constants are highly suppressed. The resulting $\dot{\mu}/\mu$ is well within atomic clock bounds.

We stress that the bound on $\dot{\mu}/\mu$ constrains the cosmological time variation of couplings, whereas static fifth-force and equivalence-principle tests probe the scalar charge of macroscopic bodies. In the present model, these are complementary observables. Both the leading $\sigma$-dependence of $m_p$ and the dominant residual nucleon coupling are controlled by the approximately universal combination $\sigma_{\pi N} + \sigma_s$. Any composition dependence is therefore expected only from subleading isospin breaking, electron couplings, and nuclear binding-energy effects, which are beyond the scope of the present paper.

\section{Mirror Pions}
\label{sec:mirror_pions}

The mirror quark masses $m_{q_D}$ are assumed to be sufficiently small ($m_{q_D} \ll \Lambda_{\text{QCD}}$) that their sigma terms, such as mirror counter parts of $\sigma_{\pi N}$ and $\sigma_s$ are negligible. This is consistent with the scale-invariant UV structure we assume. Here we may have one possible scenario to provide the origin of the mass $m_{q_D}$: It has been suggested that asymptotically safe quantum gravity with a Gaussian IR fixed point can generate small fermion masses via chiral symmetry breaking~\cite{Eichhorn:2016vvy, Daas:2021abx, deBrito:2023kow}. If such a mechanism operates in the mirror sector, it would naturally produce $m_{q_D} \ll \Lambda_{\text{QCD}}$ while preserving the classical scale invariance at high energies. However, our fifth-force prediction does not rely on this specific UV completion; any mechanism that keeps mirror quarks light suffices.

The mirror pions $\pi_D$ are the pseudo Nambu-Goldstone bosons of spontaneously broken mirror chiral symmetry, forming the mirror isospin $SU(2)$ triplet. Their mass follows from the Gell-Mann–Oakes–Renner relation:
\begin{align}
    m_{\pi_D}^2 
    = \frac{(m_{u_D}+m_{d_D}) \langle \bar{q}_D q_D \rangle}{f_{\pi_D}^2},
\label{eq:gor}
\end{align} 
where $\langle \bar{q}_D q_D \rangle \sim -\Lambda_D^3$ is the mirror chiral condensate and $f_{\pi_D} \sim \Lambda_D$ is the mirror pion decay constant. Thus, the mirror pion mass $m_{\pi_D}^2 \sim m_{q_D} \Lambda_D$. For $m_{q_D} \ll \Lambda_{\text{QCD}}$, they are light. They are not thermally coupled to the SM and can evade cosmological constraints if the mirror sector is never reheated. A detailed analysis is beyond the scope of this paper. 

Since the mirror QCD sector is completely SM-singlet and parity-invariant in the present setup, the mirror pions cannot decay into anything. They are not thermally coupled to the SM plasma due to the extreme suppression of the $\sigma$-mediated interaction ($\theta_{\sigma-G} \sim 1\,{\rm GeV}^2/M_P^2\sim 10^{-36}$), so they evade direct thermal constraints from Big Bang Nucleosynthesis or the cosmic microwave background. Their cosmological abundance depends on the initial conditions set by inflation and reheating. A minimal scenario is that the mirror sector is never reheated to significant temperatures, leaving it with negligible energy density today. Having a tiny coupling, hence a tiny cross section to the SM sector as above, mirror pions would yield a negligibly small contribution to the effective number of neutrinos, $\Delta N_{\rm eff}$, as well. 

Alternatively, if mirror pions are sufficiently heavy ($\gtrsim \text{MeV}$), they could constitute a subdominant cold dark matter component. A detailed cosmological analysis of the mirror sector's thermal history is beyond the scope of this paper, but the parameter space exists where all observational constraints are satisfied. Notably, the fifth-force prediction in the main text is independent of these details, relying only on the condition that mirror quark masses are sufficiently small to render their sigma terms (such as a mirror version of $\sigma_{\pi N}$ and $\sigma_s$) negligible.

\section{Generality of the Mirror Sector Construction}
\label{sec:mirror_generality}

The choice of a minimal $SU(3)_D$ sector with three light flavors ($u_D, d_D, s_D$) is the most predictive construction. Three flavors provide the necessary non-perturbative structures, namely the trace anomaly and chiral symmetry breaking, while mirroring the successful low-energy dynamics of QCD. A two-flavor mirror sector would lack the strange quark contribution $\sigma_s$. The resulting fifth-force prediction would then depend on the difference in gluon condensates between $N_f=3$ and $N_f=2$ QCD, a quantity that is not independently known with comparable precision, thereby reducing the predictive power of the mechanism.

Even if one were to extend the model to include a full mirror SM with heavy quarks, as long as other mirror particles are heavier than the scale of $\Lambda_{\text{QCD}}$, the present fifth-force suppression works as it is. Such extensions merely introduce additional free parameters and are not required for the screening mechanism. 

The minimal three-flavor mirror sector is therefore the most predictive and natural implementation, with the $Z_2$ breaking dominated by the quark mass differences relevant to the nucleon matrix element, while maintaining the vacuum alignment.


\bibliographystyle{apsrev4-1}
\bibliography{reference}

\end{document}